\documentclass[10pt,superscriptaddress,twocolumn,amsmath,amssymb,aps,prl,showpacs]{revtex4}
\usepackage{graphicx}
\usepackage{dcolumn}
\usepackage{bm}
\usepackage{amsmath,amssymb,mathrsfs}
\usepackage{paralist}
\usepackage[T1]{fontenc}
\usepackage{color, xcolor}
\usepackage[colorlinks,linkcolor=red,anchorcolor=blue,citecolor=blue]{hyperref}
\usepackage[caption=false,farskip=0pt,labelfont={bf}]{subfig}
\def\be{\begin{equation}}
\def\ee{\end{equation}}
\def\bea{\begin{eqnarray}}
\def\eea{\end{eqnarray}}

\begin{document}

\title{Haldane Fractional Statistics for 1D Heisenberg Spin XXX Chain}

\author{Wei-Jia Liu}
\affiliation{School of Physics and Technology, Wuhan University, Wuhan 430072, China}
\affiliation{Innovation Academy for Precision Measurement Science and Technology, Chinese Academy of Sciences, Wuhan 430071, China}

\author{Jia-Jia Luo}
\email[]{luojiajia@wipm.ac.cn}
\affiliation{Innovation Academy for Precision Measurement Science and Technology, Chinese Academy of Sciences, Wuhan 430071, China}
\affiliation{University of Chinese Academy of Sciences, Beijing 100049, China}

\author{Xi-Wen Guan}
\email[]{xwe105@wipm.ac.cn}
\affiliation{Innovation Academy for Precision Measurement Science and Technology, Chinese Academy of Sciences, Wuhan 430071, China}
\affiliation{Department of Fundamental and Theoretical Physics, Research School of Physics,
Australian National University, Canberra ACT 0200, Australia}
\affiliation{NSFC-SPTP Peng Huanwu Center for Fundamental Theory, Xian 710127, China}

\date{\today}


\begin{abstract}
   Haldane's fractional exclusion statistics (FES) describes a generalized Pauli exclusion statistics,
which can be regarded as an emergent quantum statistics induced by the intrinsic dynamical interaction. A non-mutual FES has been identified at the quantum criticality of the one-dimensional (1D) and 2D 
interacting Bose Gas [Nat. Sci. Rev. 9, nwac027 (2022)]. It is naturally asked if such a
non-mutual FES can be induced by the spin-spin interaction in the antiferromagnetic spin-1/2 XXX
chain? In this article, we first represent the Bethe ansatz equations of spin strings in terms of
the FES equations of different species. Then we show that the 1D spin XXX chain remarkably
possesses the non-mutual FES in the critical region. We observe that the equation of state in terms
of the FES gives rises to full statistical properties of the model at quantum criticality, which are
in good agreement with the results obtained from the thermodynamic Bethe ansatz (TBA) equations of the model. From the non-mutual FES, we also precisely determine the quantum scaling functions, which further agree well with the previous TBA results [Phys. Rev. B 96, 220401(R)
(2017)]. Finally, we also build up an exact mapping between the scaling functions of the Lieb-Liniger model and  the spin Heisenberg spin chain at quantum criticality. Our method provides deep insights into the critical phase of matter from quantum FES
point of view.

\end{abstract}

\maketitle
\section{Introduction}
It is well known that Bose-Einstein and Fermi-Dirac statistics form the fundamental rule of quantum statistical physics. 
Bosons with integer spin can undergo Bose-Einstein condensation (BEC), whereas fermions with half odd integer spin are not allowed to occupy a single quantum state due to the Pauli exclusion principle. 
Quantum degenerate gases of ultracold atoms open up unprecedented possibilities for the experimental study of subtle quantum statistical physics. 
However, such natural quantum statistics  of bare particles  can be generalized to the fractional one, i.e., anyonic statistics. 
In two-dimensional (2D), anyonic excitations can carry fractional charges, which obey fractional statistics \cite{leinaas1977, wilczek48, wilczek49},  also see a Nobel Lecture   \cite{Stormer:1999}.  
Haldane in 1991 further generalized it and gave a novel concept named fractional exclusion statistics (FES) that recovers the Bose-Einstein and Fermi-Dirac statistics in arbitrary dimensions  \cite{haldane1991}. 
Such a concept of the FES was further improved by Wu \cite{wuys1994, wuys1995} and other workers \cite{isakov1994, ha1994}. This concept has been successfully applied in some one-dimensional (1D) systems, including the Calogero-Sutherland model with  $1/r^2$ potential \cite{calogero1969_1, calogero1969_2, sutherland1971, wuys1995}, Lieb-Liniger model \cite{wuys1995, lieb1963}, and anyonic gases with delta-function interaction \cite{kundu1999, batchelor2006}.

FES reveals the statistical nature of quantum systems without concerning the specific form of interaction.
Such fractional statistical nature originates from the distribution of the momenta of the interacting quasiparticles in quantum systems. 
It not only generalizes basic ideas of free fermions and bosons, but also depicts the particle-hole symmetry breaking (PHSB) \cite{ha1994} induced by interaction in the strongly correlated systems, such as high-$T_c$ superconductors \cite{hashimoto2010, miller2017} and fractional quantum Hall systems \cite{zhangy2016}. Such a symmetry breaking significantly influences physical properties of the system, including equations of state \cite{bhaduri2007}, optical properties \cite{tabert2015}, dynamical evolutions \cite{balakrishnan2009}, transport properties \cite{demchenko2004} and non-Fermi-liquid behaviors \cite{kusunose1996}. However, depicting the PHSB in general interacting systems remains a challenge. The main purpose of this article is to identify the emergent FES based on the PHSB in isotropic antiferromagnetic Heisenberg spin chain near critical point.

The 1D antiferromagnetic  spin-1/2 XXX chain plays a vital role in both spin and integrable models. Bethe \cite{bethe1931} proposed a special form of wave function to solve this model in 1931, which later on was called the Bethe ansatz (BA).
Forty years later after Bethe's work, Takahashi developed the formulation of spin strings patterns \cite{takahashi1971} to analyze spinon
and bound states of spin wave excitations in this spin chain. Using Yang and Yang's thermodynamic method \cite{yang1969}, the thermodynamic equations of strings can be derived, namely, the so-called thermodynamic Bethe ansatz (TBA) equations \cite{takahashi1971}. Due to strong quantum fluctuations in one-dimensional quantum spin systems \cite{giamarchi2003}, the 1D spin XXX chain can undergo a quantum phase transition \cite{kono2015} from the Tomonaga-Luttinger liquid (TLL) \cite{haldane1980} to fully polarized ferromagnetic (FM) state. The analytical results of scaling functions near critical point were obtained in 2017 \cite{hefeng2017}, which provide a rigorous understanding of the quantum criticality of spinons. Such universal thermal and magnetic critical behaviors have been immediately confirmed in 2017 \cite{breunig2017} and 2024 \cite{channarayappa2024}.
Building on Bethe ansatz, one may investigate
mathematical models of interacting fermions associated with a new experiments on spin compounds,
electronic solids and cold atoms. However, the Bethe ansatz is usually invalid for studying higher dimensional
systems.

Nevertheless, the physical picture based on the FES can be generally true for all dimensions. The theoretical study of identifying the FES in one and two dimensional Lieb-Liniger model was carried out in \cite{zhangxb2022}. The phase diagrams of spin XXX chain and Lieb-Liniger gas present a surprising similarity, which inspire us to establish an
intrinsic connection between these two different systems. We believe that such a connection between the FES and the
quantum spin system at quantum criticality may hold true for two and three dimensions.

This paper is organized as follows. In Section \ref{sectionmodel}, we will introduce the BA equations and TBA equations of the 1D
integrable Heisenberg spin chain.
In  Section \ref{sectioncriticality}, the particle-hole symmetry breaking of spin strings will be studied to build up a general formalism of the FES in terms of spin states. Then we can get both analytical
and numerical solutions of the spin model at quantum criticality.
In Section \ref{sectionmapping}, we will analytically derive the scaling
functions in the critical region using the technique of FES, displaying universal feature  of non-interacting quasiparticles with the effective FES parameter $g_0$ within the critical region, especially the feature of spinless free fermions at low temperature. Through such feature, the mapping between
spin chain and Bose gas in one dimension can be established within the critical region. The last Section is remained for our conclusion and discussion.

\section{Antiferromagnetic Heisenberg spin-1/2 XXX chain}
\label{sectionmodel}
\subsection{Bethe Ansatz Equations and String Hypothesis}
The Hamiltonian of the 1D Heisenberg spin-1/2 chain is given by \cite{bethe1931}
\begin{equation}
    \mathcal{H}=2J\sum_{j=1}^{N}\vec{S}_j\cdot\vec{S}_{j+1}-\frac{g\mu_BH}{\hbar}M^z,
\end{equation}
where $J$ is the intrachain coupling constant, $N$ is the number of lattice site, $g$ and $\mu_B$ are the Land\'e factor and the Bohr magneton, respectively. $H$ is the external magnetic field and $M^{z} = \sum_{j=1}^N S_j^z$ is the magnetization. We set $\hbar=g\mu_B=1$ for simplicity in our calculation. Then the magnetization can be written as $M^z = N/2-M$, where $M$ is the number of down spins. The spin quasimomentum $k_j$ with $j=1,...M$ and total energy $E$ are given by Bethe ansatz (BA) equations \cite{takahashi1999, bethe1931}
\begin{align}
\mathrm{e}^{\mathrm{i}k_jN}&=-\prod_{l=1}^{M}\frac{\lambda_j-\lambda_l-\mathrm{i}}{\lambda_j-\lambda_l+\mathrm{i}}\label{BAEofrealsolution}, \\
E(\lambda_1,\cdots,\lambda_M)&=-J\sum_{j=1}^{M}\left(\frac{1}{\lambda_j^2+\frac{1}{4}}\right)+MH+E_0 \notag \\
&=\sum_{j=1}^{M}\varepsilon_1^0(\lambda_j)+E_0\label{energyofrealsolution},
\end{align}
where $\lambda_j$ is rapidity, $k_j=\pi-2\arctan{2\lambda_j}$ denotes quasimomentum   through  the relation  $ \mathrm{e}^{\mathrm{i}k_j}= \left(\lambda_j-\frac{\mathrm{i} }{2} \right)/ \left( \lambda_j+\frac{\mathrm{i} }{2}\right) $, $\varepsilon_1^0(\lambda)=-2\pi Ja_1(\lambda)+H$ with $a_1(\lambda)=\frac{1}{2\pi}\frac{1}{\lambda_j^2+1/4}$ is the energy of quasiparticle and $E_0=\frac{N}{2}(J-H)$ is the vacuum energy. For ground state, all rapidities $\lambda_j$ take real values. While for finite temperature, Takahashi found complex solutions  in thermodynamic limit, i.e., $M,N\to \infty$, $M/N$ is finite, which are called string hypothesis \cite{takahashi1971}
\begin{equation}\label{complexsolutions}
\lambda_{\alpha,j}^{n}=\lambda_{\alpha}^{n}+\frac{\mathrm{i}}{2}(n+1-2j),
\end{equation}
with $j=1,2,\cdots,n$ and $\alpha=1,2\cdots,\nu_{n}$. Here, $\lambda_{\alpha}^{n}$ and $\nu_n$ denote the real part and the number of length-$n$ strings, respectively. Under this hypothesis \cite{takahashi1971}, taking the logarithm of equation (\ref{BAEofrealsolution}), we derive 
\begin{align}\label{discreteBAEofnstrings}
\theta_n(\lambda_{\alpha}^n) = \frac{2\pi I_{\alpha}^n}{N}+\frac{1}{N}\sum_{m,\beta}\Theta_{mn}(\lambda_{\alpha}^n-\lambda_{\beta}^m),
\end{align} 
where $\theta_n(x)=\pi-2\arctan(2x/n)$, and
\begin{align}
\Theta_{mn}(x)= & \theta_{m+n}(x)+2\theta_{m+n-2}(x)+\cdots+2\theta_{|m-n|+2}(x) \notag\\
&+(1-\delta_{mn})\theta_{|m-n|}(x).
\end{align}
Here, $I_{\alpha}^n$ is an integer (half-odd integer) for $N-\nu_n$ odd (even) and satisfy
\begin{align}\label{I}
\begin{split}
|I_{\alpha}^n|\le \frac{1}{2}\left(N-1-\sum_{m=1}^{\infty}t_{nm}\nu_m\right),\\
 t_{nm}\equiv 2\mathrm{min}(n,m)-\delta_{nm}.
\end{split}
\end{align}

\begin{figure*}[t]
\includegraphics[scale=0.6]{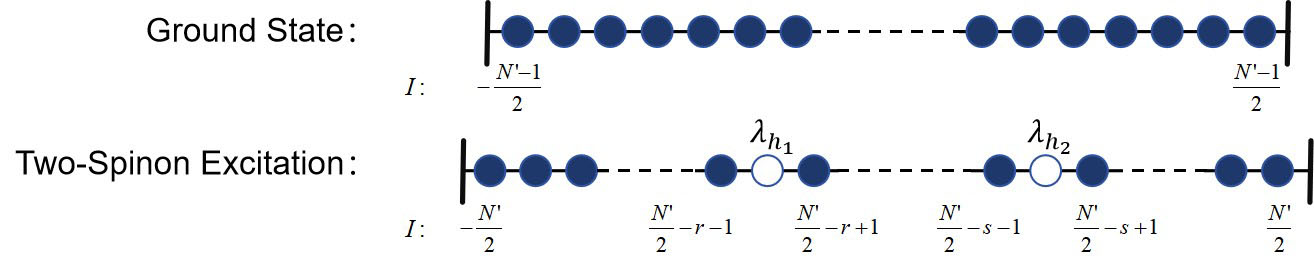}
\caption{\textbf{ Illustration of root vacancies for the  ground state configuration and the two-spinon excitation.} In the absence of external magnetic field, the vacancies at ground state are completely filled by $N/2$ magnons. In the case of one spin flipping, one more
vacancy arises with the flipped  spin, and thus two holes $\lambda_{h1},\lambda_{h2}$ are created, indicating that one magnon decomposes into two spinons with  $1/2$ spin.}\label{excitation}
\end{figure*}

It can be seen that the number of particles that can be accommodated in the $n$-string space, namely the vacancies are $N-\sum_{m=1}^{\infty}t_{nm}\nu_m$. Any set of quantum number $\{I_{\alpha}^n\}$ determines one regular solution $\left\{\lambda_{\alpha}^n\right\}$ \cite{yang1966}.
For this reason, these quantum numbers characterize full states of the system, including the ground state and excited states. Define functions $h_n(x)$ as
\begin{equation}
h_n(x)\equiv\theta_n(x)-\frac{1}{N}\sum_{m,\beta}\Theta_{mn}(x-\lambda_{\beta}^{m}).
\end{equation}
Then the quantum number $I_{\alpha}^n$ can be expressed as
\begin{equation}
\frac{2\pi I_{\alpha}^n}{N}=h_n(\lambda_{\alpha}^n).
\end{equation}
For simplicity, here we consider the  excitation of one spin flipping to understand the distributions of quantum number $I_{\alpha}^n$. We assume spin-down particles $M$ to be an odd number, in which case the ground state of spin chain has no degeneracy. At ground state with vanishing magnetic field, we have $M = N/2\equiv N'$, and quantum numbers (\ref{I}) are integers within the range $|I_{\alpha}^1|\le\frac{1}{2}(N'-1)$. Therefore the vacancies at ground state
are completely filled by all spins, whose quantum numbers are 
\begin{equation*}
I: -\frac{N'-1}{2}, -\frac{N'-3}{2}, \cdots, \frac{N'-3}{2}, \frac{N'-1}{2}.
\end{equation*}
Then we consider the excitation of one spin flipping, i.e., $ M = N' - 1$. At this excited state, the vacancies of quantum number $I_{\alpha}^1$ increase to $N' + 1$ with only $ N' - 1$ magnons distributed, rendering two holes created in the $\lambda$ space with separate $1/2$ spin, which is called two-spinon excitation, see Fig. \ref{excitation}. The distributions of quantum numbers for this excitation are
\begin{align*}
    I: &-\frac{N'}{2},\cdots, \frac{N'}{2}-r-1, \frac{N'}{2}-r+1, \cdots,\notag\\ 
    &\frac{N'}{2}-s-1,\frac{N'}{2}-s+1, \cdots, \frac{N'}{2}.
\end{align*}
Here the numbers $\frac{N'}{2}-r,\, \frac{N'}{2}-s$ denote the  positions of two spinons. In the thermodynamic limit, the distribution functions of particles and holes in the $n$-string sea are defined as $\rho_n(\lambda)$ and $\rho_n^h(\lambda)$, respectively. The numbers of particles and holes between the interval $\lambda$ and $\lambda+d\lambda$ are $N\rho_n(\lambda)d\lambda$ and $N\rho^h_n(\lambda)d\lambda$ respectively. Thus equation (\ref{discreteBAEofnstrings})
converts to
\begin{equation}\label{continueBAEofnstrings}
\rho_n^{h}(\lambda)=a_n(\lambda)-\sum_{m=1}A_{mn}\ast\rho_m(\lambda),
\end{equation}
where $a_n(\lambda)=\frac{1}{2\pi}\frac{n}{n^2/4+\lambda^2},\,a_0(\lambda)=\delta(\lambda)$ and the convolution kernel is
\begin{align}
A_{mn}(\lambda)=&a_{m+n}(\lambda)+2a_{m+n-2}(\lambda)+\cdots+2a_{|m-n|+2}(\lambda) \nonumber \\
&+a_{|m-n|}(\lambda).
\end{align}
Here we denote the convolution $\int_{-\infty}^{\infty}a(x-y)b(y)dy$ as $a\ast b(x)$ for two arbitrary functions $a(x)$ and $b(x)$.


\subsection{Thermodynamic Bethe Ansatz Equations and Thermodynamic Quantities}

\begin{figure*}[t]    
  \centering  
  \vspace{-5pt}
  \includegraphics[scale=0.55]{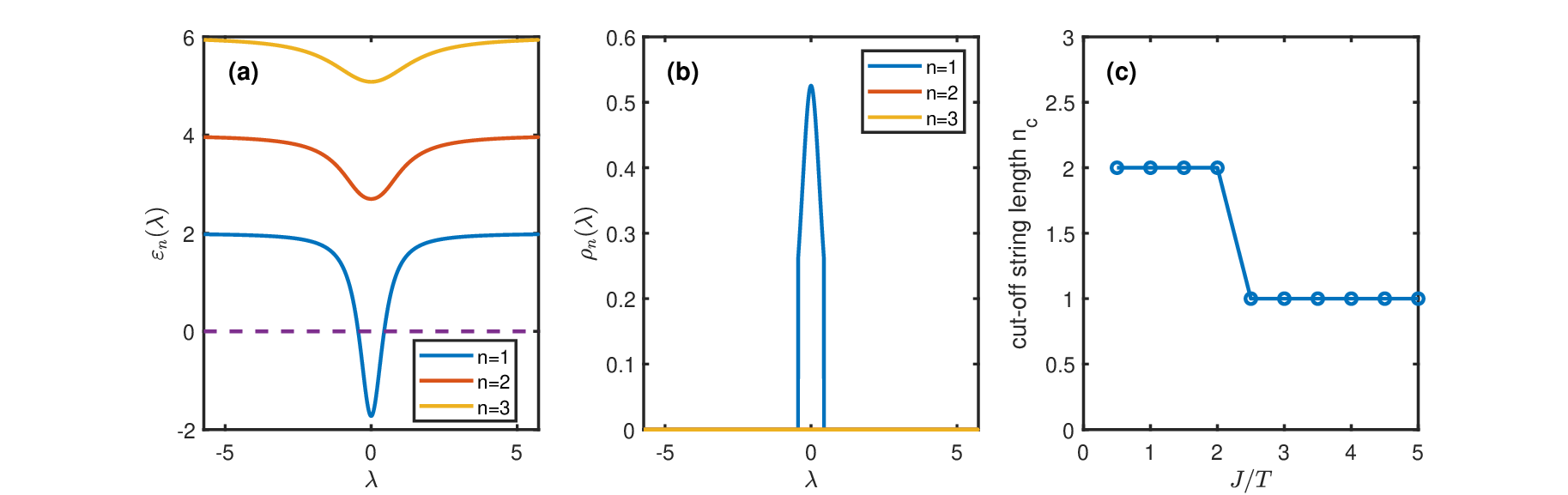} 
  \vspace{-5pt}  
\caption[Dressed energy and density distribution functions]
    {
  \textbf{(a) Dressed energies (\ref{TBA}) of spin strings at zero temperature, (b) Density distributions (\ref{continueBAEofnstrings}) of length-$n$ string  at zero temperature, and (c) Cut-off of string lengths at finite temperature.}
  (a) The dressed energies are monotonically increasing functions with rapidities $|\lambda|$. Here the parameters are $J=1,\, H = 2J$. Only for length-1 string, there exists a Fermi sea $(-Q, Q)$ satisfying $\varepsilon_n(\lambda)<0$, and for $\lambda\to\infty$, $\varepsilon_n(\lambda)$ approaches to a constant $nH$.
    (b) Only the length-1 string can contribute at zero temperature. (c) The cutoff string length $n_c$ versus interaction $J/T$ at precision of the order $10^{-6}$\cite{hefeng2017}. Here we set $H=4J$, where the system locates within the quantum critical region. It shows that the contributions from longer length strings emerge when $J/T$ become small.}
    \label{DressedEnergyPlot}
\end{figure*} 
Based on previous works, C. N. Yang and C. P. Yang developed an approach to study the thermodynamic equilibrium properties of Bose gas with delta potential \cite{yang1969}, so called Yang-Yang thermodynamic approach. Yang-Yang thermodynamics of the 1D Lieb-Liniger model has been observed in paper \cite{van2008}. Here using the  Yang-Yang thermodynamic Bethe
ansatz (TBA) method, we first define dressed energy of length-$n$ string state $\varepsilon_n=T\ln{\rho^h_n/\rho_n}=T\ln{w_n}.$ The entropy of the system  is determined by the Boltzmann relationship
\begin{align}
S=&\int dS =\sum_{n}\int \ln{\frac{\left[N(\rho_n(\lambda)+\rho^h_n(\lambda))d\lambda\right]!}{\left[N\rho_n(\lambda)d\lambda\right]!\left[N\rho^h_n(\lambda)d\lambda\right]!}}\notag\\
\approx & N\sum_n\int\Bigg[\rho_n(\lambda)\ln{\left(1+\frac{\rho_n^h(\lambda)}{\rho_n(\lambda)}\right)} \notag \\
&+\rho_n^h(\lambda)\ln{\left(1+\frac{\rho_n(\lambda)}{\rho_n^h(\lambda)}\right)}\Bigg]d\lambda.
\end{align}
The last step uses the Stirling relation $\ln{N!}\approx N\ln{N}-N$. The free energy is given by
\begin{align}
F &= E - TS, \\
E &= N\sum_n\int\varepsilon_n^0(\lambda)\rho_n(\lambda)d\lambda+E_0,
\end{align}
where $\varepsilon_n^0(\lambda)=-2\pi Ja_n(\lambda)+nH$ is the energy of length-$n$ string. Consequently, the thermodynamic equilibrium states
are determined by the minimization of Gibbs free energy  per site $f = (F-E_0)/N$, i.e., $\delta f=0$, which gives the so-called TBA equations of spin XXX chain
\begin{equation}\label{TBA}
\varepsilon_{n}^{+}=\varepsilon_n^{0}-\sum_{m}A_{mn}\ast\varepsilon_{m}^{-},
\end{equation}
where $\varepsilon_{n}^{\pm}=\pm T\ln{(1+\mathrm{e}^{\pm \varepsilon_{n}/T})}$ is the a simplistic notation, $\varepsilon_n=T\ln{\rho_n^h/\rho_n}$ is the dressed energy. Accordingly, the free energy and energy per site  is given by
\begin{align}
f&=\sum_n \int a_n(\lambda)\varepsilon_n^{-}(\lambda)\mathrm{d}\lambda, \\
e&=\sum_n \int \varepsilon_n^0(\lambda)\rho_n(\lambda)\mathrm{d\lambda},\label{e}
\end{align}
respectively. The thermodynamic quantities can be calculated through the standard thermodynamics relations
\begin{equation}
\frac{M^z}{N} = \frac{\partial f}{\partial H},\quad s = \frac{\partial f}{\partial T},\quad \chi=\frac{\partial^2 f}{\partial H^2},\quad c_H=T\frac{\partial^2 f}{\partial T^2}.
\end{equation}
According to the definition of $\rho_n(\lambda)$, the magnetization can also be calculated through
\begin{align}
m^z&=\frac{M^z}{N}=\frac{1}{2}-\frac{M}{N}\equiv\frac{1}{2}-m, \label{mz}\\
m&=\sum_nn\int \rho_n(\lambda)d\lambda,\label{m}
\end{align}
where $m=M/N$ is the number of down spins per site. Thus the susceptibility is the derivative of $m$ with respect to magnetic field, i.e., $\chi=-\partial m/\partial H.$

At zero temperature $T=0$, $\varepsilon^{\pm}_n$ become 
\begin{align}
\varepsilon^{+}_n(\lambda)=\begin{cases}
\varepsilon_n(\lambda), \quad &\text{for}\quad \varepsilon_n(\lambda)\ge0 \\
0,  \quad &\text{for}\quad \varepsilon_n(\lambda)<0 ,
\end{cases}
\\
\varepsilon^{-}_n(\lambda)=\begin{cases}
0, \quad &\text{for}\quad \varepsilon_n(\lambda)\ge0 \\
\varepsilon_n(\lambda),  \quad &\text{for}\quad \varepsilon_n(\lambda)<0 .
\end{cases}
\end{align}
From TBA equations (\ref{TBA}), it can be proved that $\varepsilon_n\ge 0$ for $n\ge 2$. Therefore, at zero temperature, only length-1 string contributes, which can also be illustrated in Fig. \ref{DressedEnergyPlot} (a) and (b). Based on this observation, at zero temperature, the TBA equations reduce to 
\begin{equation}
\varepsilon^{(0)}_1(\lambda)=-2\pi Ja_1(\lambda)+H-\int_{-Q}^Q a_2(\lambda-\mu)\varepsilon^{(0)}_1(\mu)d\mu,
\end{equation}
where $Q$ is the rapidity cut-off  determined by $\varepsilon^{(0)}_1(\pm Q)=0$. We note that $Q = 0$ or $\varepsilon^{(0)}_1(0)=0$ corresponds to critical point, which gives the saturation magnetic field $H_c=4J$ and $m^z = 1/2$.  At zero temperature, when the magnetic field increases from zero to the critical field, the system undergoes quantum phase transition from the antiferromagnetic state to the ferromagnetic state and the critical magnetic field is $H_c=4J$.

\section{FES and critical behaviour of the Heisenberg spin-1/2 XXX  chain  }
\label{sectioncriticality}
\subsection{Haldane Fractional Statistics}

In 1991 Haldane generalized the anyonic statistics to any dimensions and developed the FES theory \cite{haldane1991}.  In this formulation,
the variation of the available states, namely the number of holes in species $i$ are completely determined by the variation of existing particles $N_j$, that is
\begin{equation}\label{FES_o}
\Delta D_i=-\sum_{j}g_{ij}\Delta N_{j}.
\end{equation}
Here $D_i$ is the number of available states for a new particle adding to the species $i$, $N_j$ is the number of particles in species $j$, and $g_{ij}$ is the mutual statistics coefficient between species $i$ and $j$. Obviously, with this definition, it recovers the situations of fermions or bosons with $g_{ij}=g\delta_{ij}$, where $g=1$ or 0. 
To count the number of states with $\{N_i\}$ fixed, Haldane's FES gives
\begin{equation}
W(\{N_i\})=\prod_{i}\frac{(D_i+N_i-1)!}{N_{i}!(D_i-1)!},
\end{equation}
covering the cases of Bose-Einstein and Fermi-Dirac statistics. We notice that the equation (\ref{FES_o}) is a differential form of the FES, thus it requires an additional initial condition. We  denote $G_i$ as the bare available states, i.e., the number of available occupation in species $i$ without any particles in the system. In 1D systems $G_i=L\Delta k/(2\pi)$, where $L$ is the length, and in spin chain the length is the particle number $N$. Then the equation (\ref{FES_o}) can be converted to
\begin{equation}\label{FES_G}
D_i+\sum_{j}g_{ij}N_j=G_i,
\end{equation} 
which is the general form of FES.
Later Y. S. Wu established a general formalism of the FES in quasimomentum
space for 1D integrable Lieb-Liniger gas \cite{wuys1995}. In fact, such generalization can also be true in 2D and 3D
interacting many-body systems. This point is very important for future study of higher dimensional quantum systems from the FES
point of view.

Now we give the number of accessible states, which account for all particles and holes, and are defined as the sum of occupied particles and available states 
\begin{equation}
\tilde{D_i}(\{N_j\})=D_i(\{N_j\})+N_i=G_i+N_i-\sum_{j}g_{ij}N_j,
\end{equation}
To apply the theory of fractional statistics, our task is to identify statistical factor $g_{ij}$ of spin XXX chain. The key observation is that the strings can be treated as statistical quasiparticles within the framework of
FES theory. To construct fractional statistics, we need to convert the quasimomentum space to the
rapidity space $\lambda$ in the spin chain, which correlate to each other through Jacobi factor. Using the parameterization and string hypothesis, the total momentum of a length-$n$ string is given by
\begin{equation}
k_{\alpha}^n=\frac{1}{\mathrm{i}}\sum_j\ln{\frac{2\lambda_{\alpha,j}^{n}-i}{2\lambda_{\alpha,j}^{n}+i}}=2\operatorname{arccot} {(\frac{2}{n}\lambda_{\alpha}^{n})}.
\end{equation}
Although the solutions of strings (\ref{complexsolutions}) are complex, their total momentum is still a real number, which inspires us to use the total momentum to label length-$n$ string in $k$ space and use real part $\lambda_{\alpha}^{n}$ in $\lambda$ space. Then, the bare available states of length-$n$ string $G_{\alpha}^{n}$ can be determined by multiplying a Jacobian determinant
\begin{equation}
G_{\alpha}^{n}=\frac{N}{2\pi}\frac{\mathrm{d}k_n}{\mathrm{d}\lambda_n}\Delta \lambda_n=Na_n(\lambda_{\alpha}^{n})\Delta \lambda_n.
\end{equation}
For a solution of BA equations (\ref{BAEofrealsolution}), some of the (half-odd) integers between quantum numbers $I_{\alpha}$ and $I_{\beta}$ are occupied, while others are unoccupied.
Denote the occupied, unoccupied and total density of a length-$n$ string as $\rho_n,\rho_n^h$ and $\rho_n^t$, then we have
\begin{align*}
N\rho_n(\lambda_{\alpha}^{n})\Delta \lambda= &\text{No. of occupied particles between } \\
&(\lambda_{\alpha}^{n}-\frac{\Delta \lambda}{2}) \text{ and } (\lambda_{\alpha}^{n}+\frac{\Delta \lambda}{2})=N_{\alpha}^{n}, \\
N\rho_n^h(\lambda_{\alpha}^{n})\Delta \lambda=& \text{No. of available states between }\\
& (\lambda_{\alpha}^{n}-\frac{\Delta \lambda}{2}) \text{ and } (\lambda_{\alpha}^{n}+\frac{\Delta \lambda}{2})=D_{\alpha}^n(\{N_{\beta}^{n}\}), \\
N\rho_n^t(\lambda_{\alpha}^{n})\Delta \lambda=& \text{No. of accessible states between } \\
&(\lambda_{\alpha}^{n}-\frac{\Delta \lambda}{2}) \text{ and } (\lambda_{\alpha}^{n}+\frac{\Delta \lambda}{2})\\
=&I_n(\lambda_{\alpha}^{n}+\frac{\Delta \lambda}{2})-I_n(\lambda_{\alpha}^{n}-\frac{\Delta \lambda}{2})=\tilde{D_{\alpha}^{n}}(\{N_{\beta}^{n}\})
\end{align*} 
Evidently, the identity $\rho_n^t=\rho_n+\rho_n^h$ always holds on. 
Under this string hypothesis, the equation (\ref{FES_G}) can be written as
\begin{equation}
\tilde{D_{\alpha}^{n}}(\{N_{\beta}\})=G_{\alpha}^{n}+N_{\alpha}^{n}-\sum_{m}\sum_{\beta}g_{\alpha\beta}^{mn}N_{\beta}^{m}.
\end{equation}
In thermodynamic limit, the above becomes
\begin{equation}\label{FES}
\rho_{n}^{h}(\lambda)+\sum_{m}\int g_{mn}(\lambda-\mu)\rho_m(\mu)\mathrm{d}\mu=a_n(\lambda).
\end{equation}
Compared with equation (\ref{continueBAEofnstrings}), the mutual statistical factor $g_{mn}(\lambda-\mu)$ can be read as 
\begin{equation}
g_{mn}(\lambda-\mu)=A_{mn}(\lambda-\mu).
\end{equation}
We would like to remark that the definition of fractional exclusion statistics does not seem to give new results in
comparison with the exact Bethe ansatz description. However, it provides a quite different view from the Bethe ansatz perspective and
shed light on the thermodynamical behavior of higher dimensional quantum many-body systems. The FES form of thermodynamic equations (\ref{FES}) is more suitable to capture contributions from different types of particles, i.e., the magnon bound states. Indeed Eq. (\ref{FES}) depicts the particle-hole symmetry breaking nature of the FES.  
The significant
point is that the dynamical interaction and the quantum statistics of the interacting particles are transmutable.
Consequently, the statistical factor significantly specifies the interaction between different particles of species. 

\begin{figure*}[t]    
  \centering  
  \vspace{-5pt}
  \includegraphics[scale=0.5]{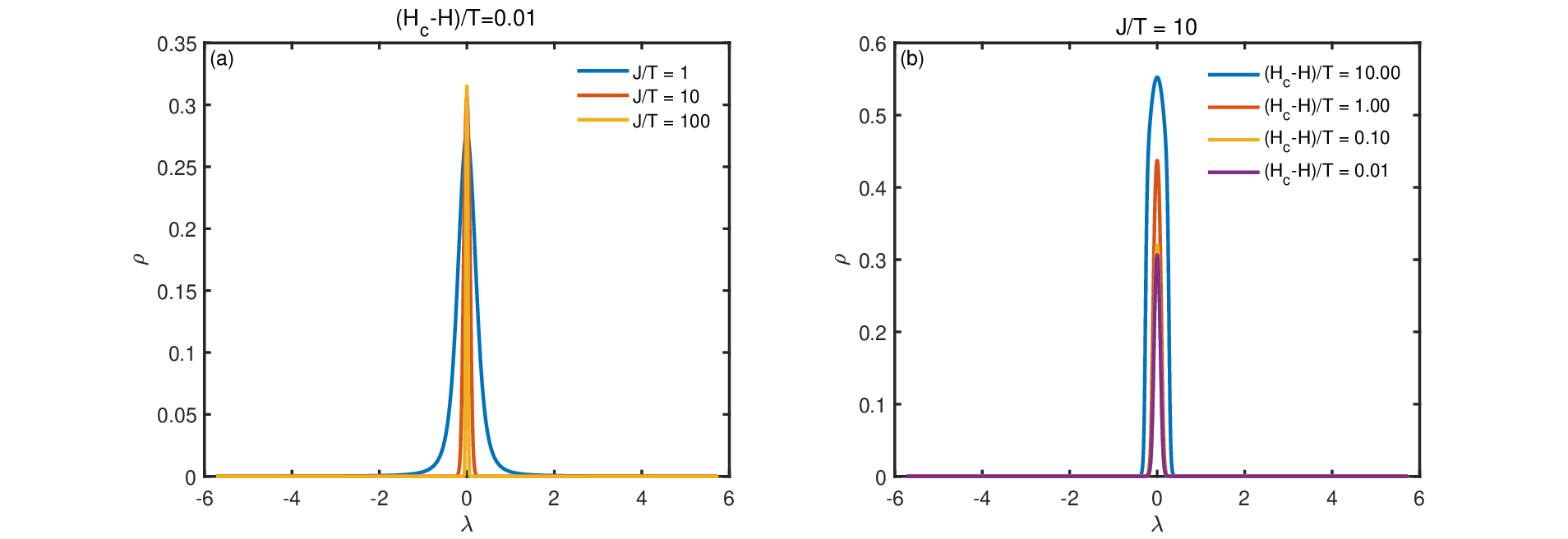} 
  \vspace{-5pt}  
\caption{\textbf{Density distributions versus rapidity at different (a) interaction strengths and (b) magnetic fields}. It can be shown that in the critical region, i.e., $|H_c-H|/T\ll1$ and $J/T\gtrsim10$, the cutoff rapidity $\lambda_c$ of the density distribution approaches to 0.} 
\label{RhoPlot}  
\end{figure*} 

In the vicinity of quantum critical point, the equal-time correlator in the spatial space has the form \cite{sachdev2011,continentino2005},
\begin{align}
	\lim_{x\rightarrow\infty}C(x,0)=Ae^{-|x|/\xi},
\end{align}
where $\xi$ is the correlation length. By taking Fourier transform of above equation, it can be deduced that the correlation length $\xi_k$ of quasimomentum space is inversely proportional to $\xi$ of spatial space \cite{zhangxb2022}. At the quantum critical point,  the correlation length approaches to infinity in spatial $\mathbf{r}$ space,  while nearly closes to zero in the quasimomentum $\mathbf{k}$ space or spin rapidity $\mathbf{\lambda}$ space, giving rise to the non-mutual FES. This deeply reflects the statistical nature of quantum criticality.

\subsection{Non-Mutual Statistics of Spin-1/2 XXX Chain}
\begin{figure*}[ht]    
  \centering  
  \vspace{-5pt}
  \includegraphics[scale=0.45]{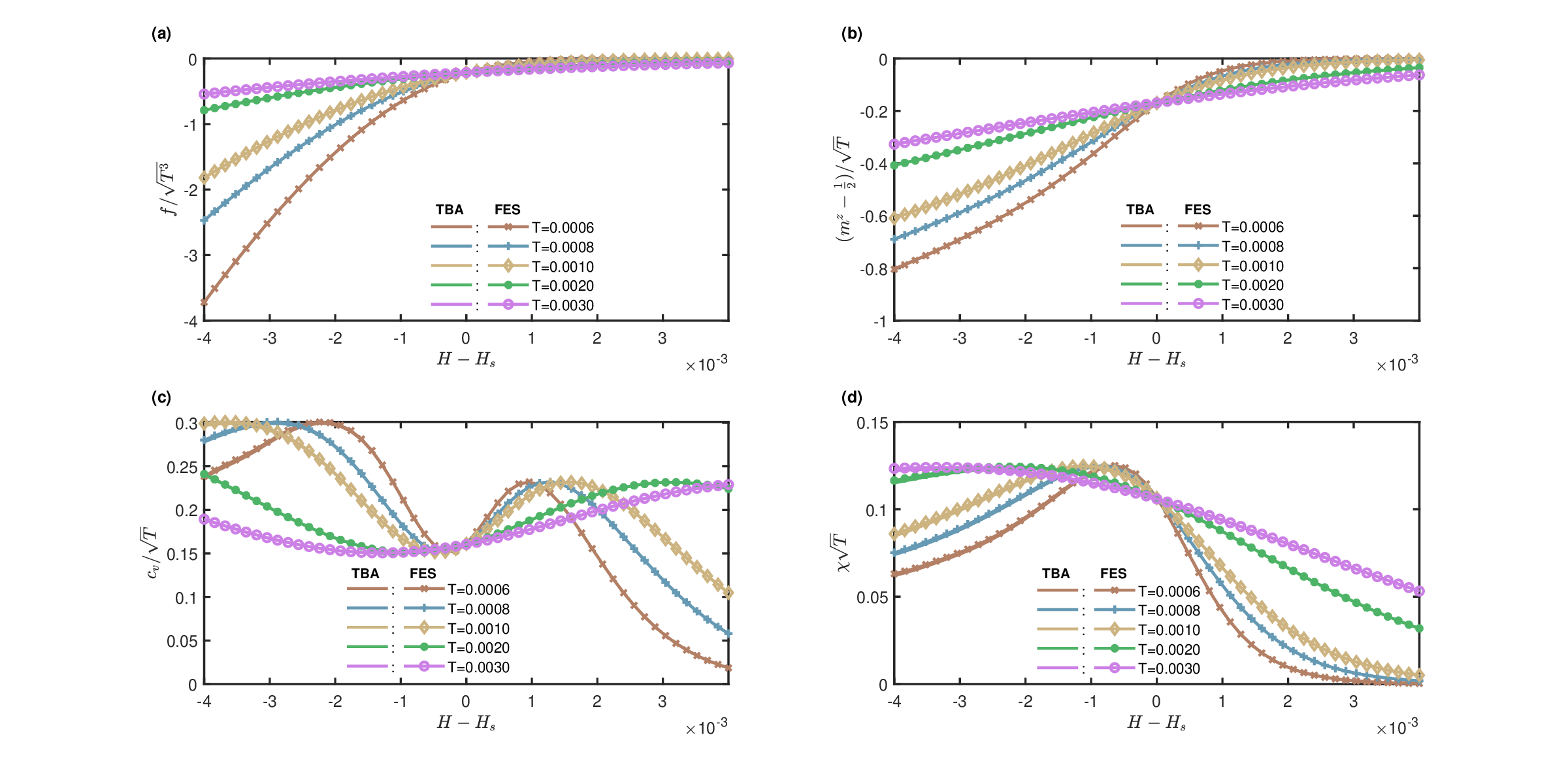} 
  \vspace{-5pt}  
  \caption{\textbf{ Scaling functions at quantum criticality: (a) Free energy, (b) Magnetization, (c) Specific heat and (d)
Susceptibility}. The results obtained from the FES are in good agreement with the numerical solutions of the TBA equations
near the critical point. Here the parameter is $J = 1$. }  
\label{scalings}  
\end{figure*} 

Based on the above analysis, a general statistics can be transformed to the non-mutual statistics in the vicinity of
the critical point, namely the statistical parameter $g_{mn}(\lambda-\mu)=g_n(\lambda)\delta_{mn}\delta(\lambda-\mu)$.  Therefore equation (\ref{FES}) can be recast as 
\begin{equation}\label{NonMutualFES}
\rho_n^h(\lambda)+g_n(\lambda)\rho_n(\lambda)=a_n(\lambda).
\end{equation}
It leads to a more concise  relation between $\delta\rho_n$ and $\delta\rho_n^h$
\begin{equation}
\delta \rho_n^h=-g_n\delta\rho_n.
\end{equation}
Our goal is to use the non-mutual statistics to derive thermodynamic quantities. The entropy per site is given by
\begin{equation}
s=\sum_n\int\big[(\rho_n+\rho_n^h)\ln{(\rho_n+\rho_n^h)} -\rho_n\ln{\rho_n}-\rho_n^h\ln{\rho_n^h}\big]\mathrm{d}\lambda.
\end{equation}
The free energy $f=e-Ts$ reaches minimum at thermodynamic equilibrium state with the condition $\delta f/\delta\rho_n=0$
\begin{align}
\delta f=&\delta e-T\delta s\nonumber\\
=&\sum_n\int\mathrm{d}\lambda[\varepsilon_n^0-T(1-g_n)\ln{(\rho_n+\rho_n^h)}+T\ln\rho_n\notag \\
&-Tg_n\ln\rho_n^h]\delta\rho_n=0,
\end{align}
which gives 
\begin{equation}
(1-g_n)\ln{(\rho_n+\rho_n^h)}-\ln\rho_n+g_n\ln\rho_n^h=\frac{\varepsilon_n^0}{T}.
\end{equation}
Let $w_n(\lambda)=\rho_n^h(\lambda)/\rho_n(\lambda)$, the above equation becomes the well known Haldane and Wu's FES equation \cite{wuys1994}
\begin{equation}\label{Wu}
(1+w_n)^{1-g_n}w_n(\lambda)^{g_n}=\mathrm{e}^{\varepsilon_n^0(\lambda)/T}.
\end{equation}
All thermodynamic quantities can be expressed with $w_n(\lambda)$, such as the per site free energy
\begin{equation}\label{FES_f}
f=-T\sum_n\int a_n(\lambda)\ln{\left(\frac{1+w_n}{w_n}\right)}\mathrm{d}\lambda.
\end{equation}

Similar analysis of the string contributions to the thermodynamics of this model can be conducted within the FES. 
At finite temperature, specifically $J/T\gtrsim10$, it can be noticed that only length-$1$ string needs to be considered in the FES equations (\ref{NonMutualFES}) and (\ref{Wu}),
\begin{align}
&\rho_{1}^{h}(\lambda)+\rho_1(\lambda)=a_1(\lambda)-{\int_{-\infty}^{\infty}} a_{2}(\lambda-\mu)\rho_1(\mu)\mathrm{d}\mu,  \label{rho1_e}\\
&\varepsilon_1(\lambda)=\varepsilon_1^0(\lambda)+T\int_{-\infty}^{\infty}a_2(\lambda-\mu)\ln{[1+\mathrm{e}^{-\varepsilon_1(\mu)/T}]}\mathrm{d}\mu. \label{epsilon1_e}
\end{align}
Furthermore, in the critical region, the magnetic field $H$ is close to the critical field, namely $|H-H_c|/T\ll 1$. The constraint about temperature and magnetic field requires that the cutoff rapidity $\lambda_c$ of the  density distribution, i.e., $\rho(\pm\lambda_c)<10^{-6}$, approaches to 0, which is elucidated in Fig. \ref{RhoPlot}. Based on above argument, we expand the integral kernel $a_2(\lambda-\mu)$ around $\mu=0$
\begin{align}
a_2(\lambda-\mu)=&\frac{1}{\pi}\Big[\frac{1}{1+\lambda^2}+\frac{2\lambda\mu}{(1+\lambda^2)^2}+\frac{(3\lambda^2-1)\mu^2}{(1+\lambda^2)^3} \notag\\
&+O(\mu^3)\Big].
\end{align}
Substituting it into equation (\ref{rho1_e}), we have
\begin{equation} 
\rho_1^{h}(\lambda)+\rho_1(\lambda)=a_1(\lambda)-\frac{m}{\pi(1+\lambda^2)}+O(\lambda_c^3),
\end{equation}
where $m=M/N$ is the number of down spins determined by the solutions of FES equations. Now, we will transform it to the form of non-mutual statistics (\ref{NonMutualFES}). Arrange above equation as
\begin{align}
a_1(\lambda)=&\frac{1}{1-\frac{m}{\pi(1+\lambda^2)a_1(\lambda)}}(\rho_1^h+\rho_1) \notag\\
\approx&\rho_1^h(\lambda)+\Bigg\{\left[\frac{m}{2-m}+\frac{6m\lambda^2}{(2-m)^2}\right]w_1(\lambda)\notag\\
&+\frac{2}{2-m}+\frac{6m\lambda^2}{(2-m)^2}\Bigg\}\rho_1(\lambda).
\end{align}
Compared with the expression of non-mutual FES (\ref{NonMutualFES}), we can read out the non-mutual statistics factor 
\begin{align}\label{g}
g_1(\lambda)&=\left[\frac{m}{2-m}+\frac{6m\lambda^2}{(2-m)^2}\right]w_1(\lambda) +\frac{2}{2-m}+\frac{6m\lambda^2}{(2-m)^2}\notag\\
&=1+\frac{m}{2-m}(1+\frac{6\lambda^2}{2-m})(1+w_1(\lambda)) \notag \\
&\approx 1+\frac{m}{2}(1+\frac{m}{2})(1+w_1(\lambda)).
\end{align}
Then, the non-mutual statistical factor $g_1(\lambda)$ can be obtained using numerical solutions, where $w_1(\lambda)$ is related to $g_1(\lambda)$ of FES equation (\ref{Wu})
\begin{equation}\label{Wu_1}
(1+w_1)^{1-g_1}w_1(\lambda)^{g_1}=\mathrm{e}^{\varepsilon_1^0(\lambda)/T}.
\end{equation}
Here $\varepsilon_1^{0}(\lambda)=-2\pi Ja_1(\lambda)+H$ and the density  of down spins $m$ is determined by the integral over (\ref{m}) with $n=1$
\begin{equation}\label{m_1}
m = \int\rho_1(\lambda)\mathrm{d}\lambda = \int\frac{a_1(\lambda)}{w_1(\lambda)+g_1(\lambda)}\mathrm{d}\lambda.
\end{equation}
Theoretically, we can iteratively solve these three equations (\ref{g}) (\ref{Wu_1}) (\ref{m_1}) to obtain all the thermal properties around the critical point. Subsequently, all the thermodynamic quantities of the spin XXX chain can be obtained from the FES method.

\subsection{The Numerical Solutions of the FES in Quantum Critical Region}
\begin{figure*}[t]    
  \centering  
  \vspace{-5pt}
  \includegraphics[scale=0.55]{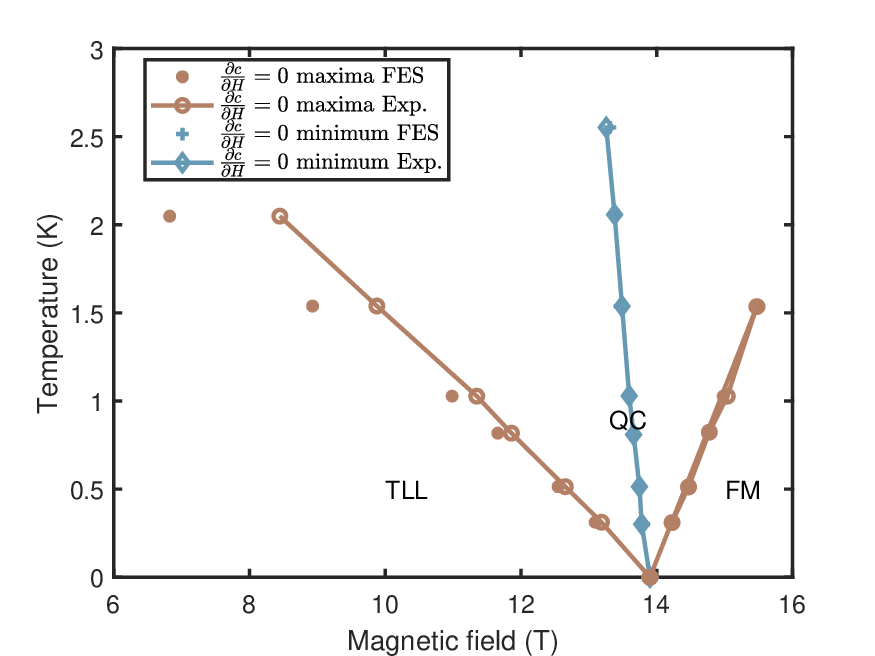} 
  \vspace{-5pt}  
\caption{\textbf{Critical temperatures of spin XXX chain around quantum critical point}. Here we compare the numerical results of FES with the experimental results of CuPzN \cite{breunig2017}. The experimental parameters of the material are: $2J/k_B = 10.6\,\mathrm{K},\, g=2.27$. To the left (right) of quantum critical point, the system lies in TLL (FM) phase. This figure exhibits the characteristics of quantum phase transition. The boundaries of the quantum critical region are determined by the double peaks of specific heat.  It can be found that the lower the temperature or the stronger the magnetic field, the better the non-mutual FES works.} 
\label{PhaseDiagram}  
\end{figure*} 
In this part, we show that the quantum criticality of spin XXX chain can be solved through non-mutual fractional statistics. To simplify notation, the subscripts will be omitted in remaining context except $a_1(\lambda)$ and $a_2(\lambda)$.
We note that
$g(\lambda)$, $w(\lambda)$ and $m$ can be obtained by solving the three coupled equations (\ref{g}) (\ref{Wu_1}) (\ref{m_1}) iteratively. The relation between the magnetization per site $m^z$ and the number of down spin $m$ is given by (\ref{mz}). The behavior of $m^z$ as function of $H$ is plotted in Fig. \ref{scalings}(b). Then the free energy per site can be calculated easily and plotted in Fig. \ref{scalings}(a)
\begin{equation}
f=-T\int a_1(\lambda)\ln{\left(1+\frac{1}{w(\lambda)}\right)}\mathrm{d}\lambda .
\end{equation}

From the equations (\ref{e}) and (\ref{m_1}), the specific heat and susceptibility are respectively derivatives of energy and magnetization 
with respect to temperature and magnetic field
\begin{align}\label{c_H}
c_v&=\frac{\partial e}{\partial T}=-\int \frac{\varepsilon_0(\lambda)a_1(\lambda)}{(w+g)^2}\left(\frac{\partial w}{\partial T}+\frac{\partial g}{\partial T}\right)\mathrm{d}\lambda ,\\
\chi&=-\frac{\partial m}{\partial H}=\int \frac{a_1(\lambda)}{(w+g)^2}\left(\frac{\partial w}{\partial H}+\frac{\partial g}{\partial H}\right)\mathrm{d}\lambda.
\end{align}
Taking specific heat as an example, the derivative of non-mutual statistical factor also needs to be calculated by solving the three coupled equations, which are given by the partial derivative of the equations (\ref{g}) (\ref{Wu_1}) (\ref{m_1}) with respect to $T$
\begin{align}
\frac{\partial g}{\partial T} = \frac{1}{2}(1+m)(1+w)\frac{\partial m}{\partial T}+\frac{m}{2}(\frac{m}{2}+1)\frac{\partial w}{\partial T},\\
\frac{\partial g}{\partial T}\ln{\left(1+\frac{1}{w}\right)}+\left(\frac{g-1}{1+w}-\frac{g}{w}\right)\frac{\partial w}{\partial T}=\frac{\varepsilon_0}{T^2},\\
\frac{\partial m}{\partial T}=-\int \frac{a_1(\lambda)}{(w+g)^2}\left(\frac{\partial w}{\partial T}+\frac{\partial g}{\partial T}\right)d\lambda.
\end{align}
Using above procedure, all thermodynamic quantities can be obtained. The specific heat and susceptibility are presented in Fig. \ref{scalings}(c) and \ref{scalings}(d). From the Fig. \ref{scalings}(c) and \ref{scalings}(d), we observe that there exists two maxima and one minimum in specific heat and one maximum in susceptibility. The maxima of the specific heat elegantly mark the crossover critical
temperatures fanning out from the critical point. The three extreme values of specific heat merge around the
quantum transition, corresponding to the critical point of the 1D spin XXX chain. The phase diagram
of spin XXX chain around quantum critical point is shown in Fig. \ref{PhaseDiagram}, where we compare the numerical results of non-mutual FES with the experimental results of CuPzN \cite{breunig2017}. It can be seen that as the magnetic field increases, the system exhibits quantum phase transition from Tomonaga-Luttinger liquid (TLL) state to the ferromagnetic (FM) phase, with the quantum critical region fanning out from the critical point. These figures demonstrate that the non-mutual FES is extremely effective at low temperatures within the quantum critical region.

\section{INTRINSIC CONNECTION BETWEEN THE SPIN-1/2 XXX CHAIN AND THE LIEB-LINIGER
GAS AT QUANTUM CRITICALITY}
\label{sectionmapping}
\subsection{Rapidity-independent non-mutual statistics and the free Fermion nature of quantum criticality }

\begin{figure*}[t]    
	\centering  
	\includegraphics[scale=0.5]{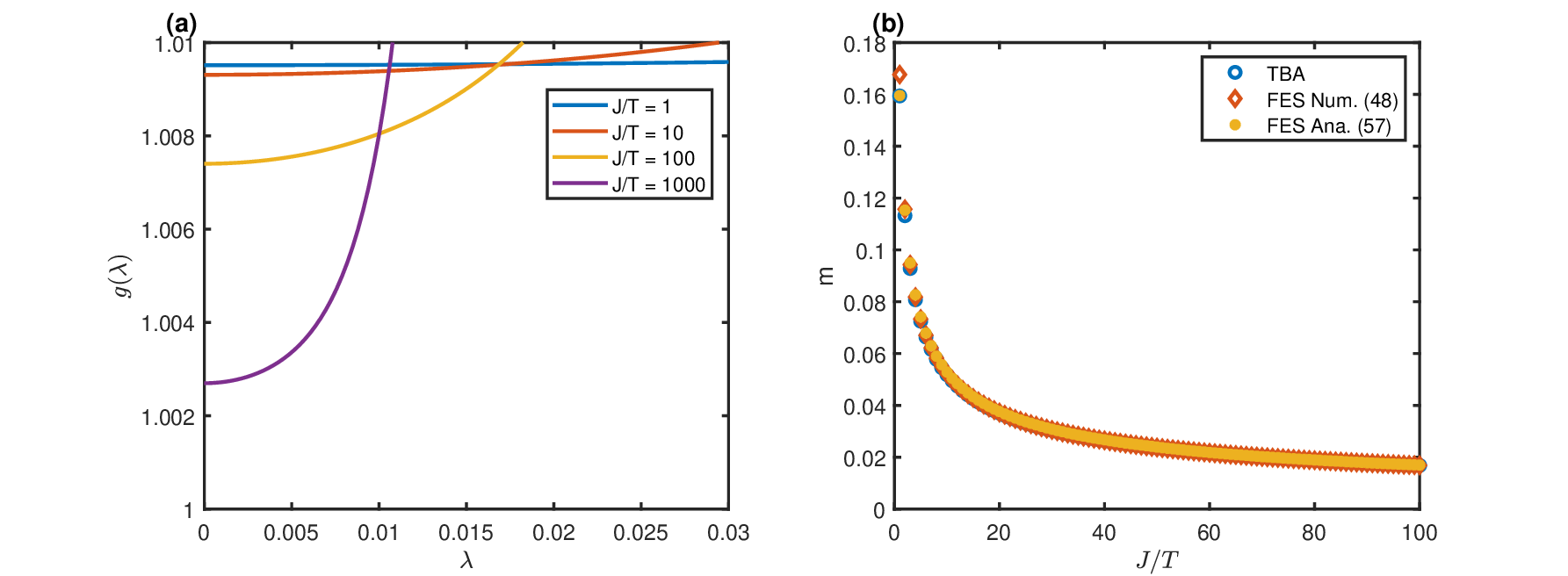}    
		\caption{\textbf{(a) Non-mutual statistical parameters and (b) Magnetization versus interaction strengths}. (a) It can be shown that as the interaction strength decreases, or equivalently as the temperature increases, the non-mutual statistical parameter $g(\lambda)$ varies more smoothly with rapidity $\lambda$ and moves further away from 1. Moreover, larger $J$ means smaller  rapidity cutoff $\lambda_c$ (see Fig. \ref{RhoPlot} (a)), so only the small $\lambda$ needs to be considered, implying the approximation $g(\lambda)\approx g(\lambda=0)$ is reasonable. (b) The magnetization as function of interaction strength for the method of TBA, rapidity-independent FES with numerical expression (\ref{m_1}) and analytic result (\ref{m_analy}). The results of TBA and FES differ greatly at $J/T=1$, which originates from the emergence of length-2 string. Both subfigures show that the $\lambda$-independent non-mutual statistical parameters are valid for $J/T>1$.}   \label{gPlot} 
\end{figure*} 
The study of the dilute magnons is very difficult \cite{schulz1980, shaginyan2017, jeong2015, maeda2007}. Here we observe that the non-mutual statistical parameter $g(\lambda)$ can be simply expressed as $g_0 = g(\lambda=0)$ in the critical region, which is characterized by Fig \ref{gPlot}. And we present a perspective of the non-mutual fractional
statistics on the criticality of spinons, i.e., the decomposed magnons.

We notice that the thermodynamic quantities can be obtained by using FES equation (\ref{Wu}) with the statistical parameter $g_0$. Based on Fig. \ref{gPlot}(a), the left-hand side of equation (\ref{Wu}) can be expanded near $g_0\approx1$ for $J/T>1$
\begin{align}
&\left \{1+(1-g_0)\ln{\left[1+\frac{1}{w(\lambda)}\right]}+O((1-g_0)^2)\right\}w(\lambda) \notag\\
=& \mathrm{e}^{\varepsilon_0(\lambda)/T},
\end{align}
with the leading solution
$ w(\lambda)= \mathrm{e}^{\varepsilon_0(\lambda)/T}$.
The density distribution of non-mutual fractional statistics is also simplified as 
\begin{equation}
\rho(\lambda)=\frac{a_1(\lambda)}{\mathrm{e}^{\varepsilon_0(\lambda)/T}+g_0}\approx \frac{2}{\pi}\frac{1}{\mathrm{e}^{[16J\lambda^2-(4J-H)]/T}+g_0},
\end{equation}
where $a_n(\lambda)\approx \frac{2}{n\pi}(1-\frac{4}{n^2}\lambda^2+\cdots)$. It's exactly the Fermi-Dirac statistics of the free fermion when $g_0=1$, where $16J\lambda^2$ is the kinetic energy and $4J-H$ is the chemical potential. The magnetization is given by
\begin{align}\label{m_analy}
m&=\int_{-\infty}^{+\infty}\frac{2}{\pi}\frac{1}{\mathrm{e}^{[16J\lambda^2-(4J-H)]/T}+g_0}d\lambda \notag \\
&=-\frac{T^{\frac{1}{2}}}{2g_0\sqrt{\pi J}}\mathrm{Li}_{\frac{1}{2}}(-g_0\mathrm{e}^{\frac{4J-H}{T}}),
\end{align} 
where $\mathrm{Li}_{k}(x)=\sum_{n=1}^{\infty}\frac{x^n}{n^k}$ is the polylogarithm function. The derivative of $m$ gives susceptibility
\begin{equation}\label{chi_analy}
\chi = -\frac{\partial m}{\partial H}=-\frac{1}{2g_0\sqrt{\pi JT}}\mathrm{Li_{-\frac{1}{2}}}(-g_0\mathrm{e}^{\frac{\Delta}{T}}),
\end{equation}
where we denote $\Delta = 4J-H$. The energy per site is 
\begin{align}\label{e_analy}
e &= \int \varepsilon_0(\lambda)\rho(\lambda)d\lambda\approx\frac{2}{\pi}\int \frac{16J\lambda^2-(4J-H)}{\mathrm{e}^{[16J\lambda^2-(4J-H)]/T}+g_0}d\lambda \notag \\
&=\sqrt{\frac{T}{\pi J}}\left[\frac{T}{4g_0}\mathrm{Li_{\frac{3}{2}}}(-g_0\mathrm{e}^{\frac{\Delta}{T}})+\frac{\Delta}{2g_0}\mathrm{Li_{\frac{1}{2}}}(-g_0\mathrm{e}^{\frac{\Delta}{T}})\right].
\end{align} 
Then the specific heat is given by 
\begin{align}\label{c_analy}
&c_v = \frac{\partial e}{\partial T} = \sqrt{\frac{T}{\pi J}}\Bigg[-\frac{3}{8g_0} \mathrm{Li_{\frac{3}{2}}}(-g_0\mathrm{e}^{\frac{\Delta}{T}}) \notag \\
&+\frac{1}{2g_0}\left(\frac{\Delta}{T}\right)\mathrm{Li_{\frac{1}{2}}}(-g_0\mathrm{e}^{\frac{\Delta}{T}})-\frac{1}{2g_0}\left(\frac{\Delta}{T}\right)^2\mathrm{Li_{-\frac{1}{2}}}(-g_0\mathrm{e}^{\frac{\Delta}{T}})\Bigg].
\end{align}

\begin{figure*}[tbp] 
\centering
    \includegraphics[scale = 0.45]{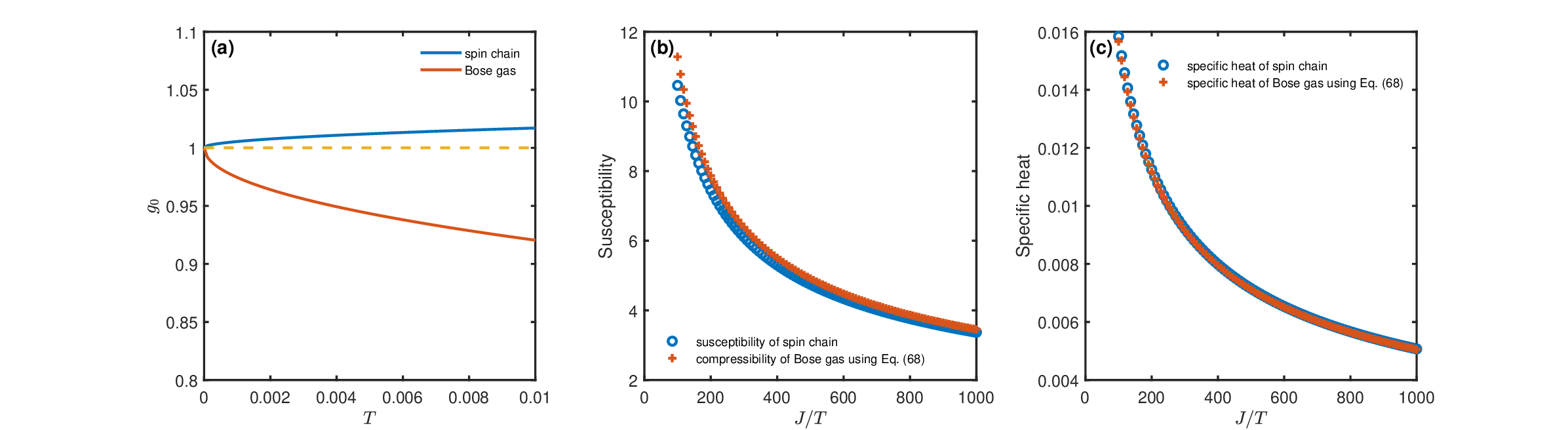}
    \caption
    {\textbf{(a) Statistical parameters $g_0$ versus temperature for spin chain and Bose gas at the critical point,  (b) Susceptibility and (c) Specific heat of spin chain and Bose gas at large interaction strength.} (a) The dash line denotes the statistical parameter of the free fermions. It indicates that at the quantum critical point, the details of  interaction in these two models become irrelevant, and their statistical properties are identical to each other, consistent with the feature of free fermions. In the critical region, as the temperature increases, the statistical properties of the quasiparticles in these two systems gradually deviate from the free fermion behavior. Here the parameters are $J=c=1$, $H=4J$ and $\mu=0$. (b), (c) Within the critical region, the spin chain and the Bose gas can still correspond to the spinless ideal particles  with effective $g_0$. Based on the mapping relation (\ref{mapping}), we plot the behavior of thermodynamic quantities of these two models.
    These two subfigures show that the mapping (\ref{mapping}) is more accurate  when $J/T\gg 1$, corresponding precisely to quantum critical region. Here the parameters are set to $H_c-H=0.01T$ to ensure these two systems closer to their critical points.}    \label{MappingPlot}
\end{figure*}

At the critical point $T=0$, $g_0=1$, the statistical behaviors of quasiparticles exhibit Fermi-Dirac statistics. The analytical solutions of thermodynamic quantities (\ref{m_analy}) (\ref{chi_analy}) (\ref{c_analy}) are consistent with the results given in \cite{hefeng2017} in the limit of $g_0=1$. In Fig. \ref{gPlot}(b), we compare the results of rapidity-independent FES, numerical expression (\ref{m_1}) and analytic expression (\ref{m_analy}) with the TBA calculations. It can be observed that the image of non-interacting quasiparticles with statistical parameter $g_0$ is reasonable in the critical region, as long as only length-1 string makes  major  contribution, i.e., $J/T>2$. And the analytical expression (\ref{m_analy}) is effective in the whole interaction range. From the equations (\ref{m_analy}) (\ref{chi_analy}) (\ref{e_analy}) and (\ref{c_analy}), it can be found that these thermodynamic quantities can be equivalent to those of spinless free fermions \cite{hefeng2017}, if the interaction strength and magnetic field can be expressed by the effective interaction strength and effective magnetic field,  respectively
\begin{equation}\label{correction}
J^{*} = Jg_0^2, \quad H_c-H^{*}=H_c-H+T\ln{g_0}.
\end{equation}
The above expressions reveal that the detail of spin-spin interaction doesn't matter within the critical region.  At low temperatures, the spin XXX chain can still be mapped onto the spinless free fermions, with the temperature only inducing corrections to the interaction strength and magnetic field. Correspondingly, The interaction plays the role of mass, and the external field corresponds to the chemical potential of the spinless free fermions.

\subsection{Mapping to the Lieb-Liniger Gas}
We notice that, the phase diagrams of the spin-1/2 XXX chain and $\delta$-function interacting Bose gases exhibit a striking consistency in the critical region, which inspires us to establish a mapping between these two systems from the FES perspective. 

In the critical region and with interaction strength of $c/\sqrt{T}\gtrsim 10$, 1D Lieb-Liniger Bose Gas is equivalent to 1D non-interacting quasiparticles with a constant statistical parameter $g_0$ \cite{zhangxb2022}. It can be observed that, similar to the spin-1/2 XXX chain, $g_0$ in Bose gas is also close to 1, but $g_0 < 1$. In Fig. \ref{MappingPlot}(a), we plot the behavior of  statistical parameters $g_S$ and $g_B$ of spin chain and Bose gas as function of temperature, in comparison with the constant statistical parameter of the free fermions which is denoted by the dash line. It shows that $g_S=g_B=1$ at zero temperature, implying both systems can be mapped onto the spinless free fermions at the  critical point. It can also be observed that as the temperature increases, $g_S$ and $g_B$ gradually diverge, indicating that the spin XXX chain and Bose gas can correspond to non-interacting quasiparticles with different statistical parameters $g_0$. At finite temperature within the critical region, the effect of $g_0$ is merely to modify the expressions of effective interaction strength and effective external field. Therefore, the correspondence between these two systems can still be established using the image of spinless free fermions with effective parameters. Similar to the magnetization (\ref{m_analy}) of spin chain, the dimensionless density of the Bose gas is written as
\begin{equation}\label{n}
\tilde{n} = \frac{n}{c} = -\sqrt{\frac{m_{BG}k_BT}{2\pi\hbar^2c^2g_B^2}}\mathrm{Li}_{\frac{1}{2}}(-g_B\mathrm{e}^{\mu_{BG}/k_BT}).
\end{equation}
On the other hand we rewrite the equation (\ref{m_analy}) with the omitted physical constants
\begin{equation}\label{m_dimen}
m = -\frac{(k_BT)^{1/2}}{2\sqrt{\pi\hbar^2Jg_S^2}}\mathrm{Li}_{\frac{1}{2}}(-g_S\mathrm{e}^{\mu_Bg(H_c-H)/k_BT}).
\end{equation}
By using the image of the spinless free fermions, we can naturally construct the mapping between spin chain and Bose Gas within the critical region. For 1D spinless free fermion system with length $L$, the spectrum is $\varepsilon=\frac{\hbar^2k^2}{2m_{FG}}$ and the density of the state is
\begin{align}
D(\varepsilon)d\varepsilon = \frac{2L}{2\pi}dk=\frac{L}{\pi\hbar}\sqrt{\frac{m_{FG}}{2}}\varepsilon^{-\frac{1}{2}}d\varepsilon.
\end{align}
Thus the density number per length is
\begin{align}\label{density_fermion}
\frac{N_{FG}}{L}&=\frac{1}{L}\int_0^{\infty}D(\varepsilon)f(\varepsilon)d\varepsilon \notag\\
&=-\sqrt{\frac{m_{FG}k_BT}{2\pi\hbar^2}}\mathrm{Li}_{\frac{1}{2}}(-\mathrm{e}^{\mu_{FG}/k_BT}),
\end{align}
where $f(\varepsilon)=1/[\mathrm{e}^{(\varepsilon-\mu)/k_BT}+1]$ is the Fermi distribution function.  By observing equations (\ref{n}) (\ref{m_dimen}) and (\ref{density_fermion}), we can establish the mapping between spin chain and Bose gas
\begin{align}
&A\sqrt{\frac{m_{FG}}{2\pi}}=\sqrt{\frac{m_{BG}}{2\pi c^2g_B^2}}=\frac{1}{2\sqrt{\pi Jg_S^2}}, \\
&\mu_{FG}= \mu_{BG}+k_BT\ln{g_B}=\mu_Bg(H_c-H)+k_BT\ln{g_S} ,
\end{align}
where $A$ is a constant with the same dimension as $1/c$ based on dimensional analysis. Considering the critical chemical potential $\mu_c$ of Bose gas satisfies $\mu_c=0$, we rewrite the above equations as
\begin{equation}\label{mapping}
\frac{g_S}{g_B}\sqrt{J} = c,\quad H_c-H+T\ln{\frac{g_S}{g_B}}=\mu,
\end{equation}
where the parameters are set to $\hbar=k_B=2m_{BG}=\mu_Bg=1$ in our calculation. The equation (\ref{mapping}) is an exact mapping between spin-1/2 XXX chain and $\delta$-function interacting Bose gas in one dimension within the quantum critical region. Using this  mapping, we can obtain the interaction strength and chemical potential of the Bose gas. Then we can plot the compressibility $\tilde{\kappa}=\kappa/c$ and specific heat $\tilde{c_v}=c_v/c$ of Bose gas as well as the susceptibility $\chi$ and specific heat $c_H$ of spin chain, see  Fig. \ref{MappingPlot} (b) and (c). From subfigures (b)((c)), we observe that that the compressibility (specific heat) of the Bose gas obtained from this mapping show an excellent  agreement with that of the spin Heisenberg chain  within the quantum critical region at low temperature. This agreements strongly support the validity of the mapping Eq. (\ref{mapping}) in the vicinity of quantum critical point ($J/T\gg 1$).  Above analysis is reasonable based on the following physical discussions. Firstly, the effect of the magnetic field and the chemical potential is opposite in these two systems that the former reduces the number of downward spins while the latter increases the number of bosons, which also accounts for the difference by the minus sign ($H_c-H$ and $\mu-\mu_c$). This minus sign also shows the difference in the transition process that the spin XXX chain transforms from Tomonaga-Luttinger liquid to ferromagnetic phase, while the Bose gas transforms from vacuum state to TLL phase. Secondly, the spin coupling constant $J$ of the spin XXX chain corresponds to interaction constant $c$ of the Bose gas, both of which represent the two-body interaction. Although the spin-spin interaction and boson-boson interaction determine  the fractional FES parameters quite differently, as shown in Fig. \ref{MappingPlot} (a), their critical behaviors can be uniformly characterized by the statistical parameters. This indicates that in the quantum critical region, the microscopic details of interactions are of less importance, reflecting the universal characteristics of quantum phase transitions.

\section{CONCLUSION AND DISCUSSION}
In this paper, we have identified the emergent FES in the 1D spin-1/2 XXX chain, and observed universal feature of
the non-mutual statistics in the vicinity of the critical point. With the help of definition of Haldane fractional statistics, we
have found that the quasiparticles in critical region exhibit the FES behavior. Accordingly, the BA equations describe
the particle-hole symmetry breaking, which remarkably depicts the FES. This originates from that the correlation
length in momentum space tends to zero at phase transition. Moreover,  with the help of the Bethe ansatz equations, we have
obtained non-mutual statistics, indicating the transmission between quantum statistics and dynamical interaction in
1D quantum systems. Using the FES equations, we have obtained various thermodynamic quantities, consistent
with the results from TBA calculations. Furthermore, the quasiparticles in both the spin chain and the
Bose gas systems exhibit the characteristics of 1D non-interacting quasiparticles with the  non-mutual statistical parameter $g_0$ in the critical region. In  general, the statistical parameter captures the statistics  of the constituent  particles and the dynamical interaction between two particles. So the mapping between these two models at low temperature within the critical region has been established through the feature of free fermions.

We would like to mention that the FES  captures the statistical nature of interacting quasiparticles which results from  intrinsic bare statistical interaction and dynamical interaction, naturally generalizing the Bose-Einstein and Fermi-Dirac statistics. 
As such, the FES represents a universal statistical behavior which  is not limited to integrable models. 
In particular, for  the same universality class of quantum criticality, i.e. same critical exponents, all thermodynamical properties are  insensitive to the microscopic details of the systems  and can be cast into the universal scaling forms. 
Such a universal phenomenon  can be naturally described by ideal particles that obey the FES in 1D and higher dimensions.
 In this regard,  quantum integrable models provides a powerful theoretical framework to identify the fractional statistics in a whole parameter space.
 For example, the FES description of the critical scaling functions in 1D and higher dimensional Bose gases at the quantum criticality was elegantly confirmed in the recent paper \cite{zhangxb2022}. 
Based on our results, one naturally expects a further study of the FES at the quantum criticality of isotropic antiferromagnetic
Heisenberg model in two dimension. 
One may also establish the dimension-independent mapping between higher dimensional spin system and the interacting Bose gas at quantum criticality. 
This opens a study on an intrinsic connection between
the Bose gas in two-dimensional systems  \cite{zhangxb2022} and the FES behavior of the two-dimensional Heisenberg model.

\section{ACKNOWLEDGEMENTS}
This work was supported by the NSFC key grants No. 12134015, and No. 92365202 and the National Key R\&D Program of China under grants No. 2022YFA1404102.
 WJL thanks the Innovation Academy for Precision Measurement Science and Technology, Chinese Academy
of Sciences for kind hospitality.

\end{document}